# Imaging and monitoring the Reykjanes supercritical geothermal reservoir in Iceland with time-lapse CSEM and MT measurements

M. DARNET[1], N. COPPO[1], P. WAWRZYNIAK[1], S. NIELSSON[2],

G.O. FRIDLEIFSSON[3], E. SCHILL[4]

[1]BRGM, France, [2]ISOR, Iceland, [3]HS-ORKA, Iceland, [4]KIT, Germany

m.darnet@brgm.fr

**Keywords:** Controlled-Source Electro-Magnetic, Magneto-Telluric, monitoring, time-lapse, geothermal reservoir, MT, CSEM

## ABSTRACT

We have investigated the benefits and drawbacks of active EM surveying (Controlled-Source EM or CSEM) for monitoring geothermal reservoirs in the presence of strong industrial noise with an actual time-lapse survey over the Reykjanes geothermal field in Iceland before and after the thermal stimulation of the supercritical RN-15/IDDP-2 geothermal well.

It showed that a high CSEM survey repeatability can be achieved with electric field measurements (within a few percent) but that time-lapse MT survey is a challenging task because of the high level of cultural noise in this industrialized environment. To assess the quality of our CSEM dataset, we inverted the data and confronted the resulting resistivity model with the resistivity logged in the RN-15/IDDP-2 well. We obtained a good match up to 2-3km depth, i.e. enough to image the caprock and the liquid-dominated reservoir but not deep enough to image the reservoir in supercritical conditions. To obtain such an image, we had to jointly invert legacy MT data with our CSEM data.

On the monitoring aspects, the analysis of changes in electric fields did not allow to identify any CSEM signal related to the thermal stimulation of the RN-15/IDDP-2 well. One possible explanation is the weakness of the time-lapse CSEM signal compared the achieved CSEM survey repeatability as a result of a limited resistivity change over a limited volume within the reservoir.

## 1. INTRODUCTION

Surface geophysical monitoring techniques are important tools for geothermal reservoir management as they provide unique information on the reservoir development away from boreholes. For magmatic environments, electromagnetic (EM) methods are attractive monitoring tools as they allow to characterize the reservoir and hence potentially monitor changes related to fluid injection/production. Indeed, the electrical resistivity of reservoir rocks is highly dependent on the volume, chemistry and phase of the in-situ geothermal brine (e.g. liquid, vapor, supercritical). Passive EM techniques (e.g. magnetotellurics or MT) are traditionally used for geothermal exploration and a few recent studies have demonstrated its potential for monitoring reservoir development. One of the main challenges is though the presence of cultural noise and/or variability of the Earth magnetic field that can obfuscate the EM signals of interest.

In the framework of the H2020-DEEPEGS project, we have investigated the benefits and drawbacks of active EM surveying (Controlled-Source EM or CSEM) to tackle this challenge, first with a synthetic study and subsequently with an actual time-lapse survey acquired in 2016 and 2017 over the Reykjanes geothermal field in Iceland before (baseline) and after (monitor) the thermal stimulation of the supercritical RN-15/IDDP-2 geothermal well.

## 2. CSEM AND MT SENSITIVITY STUDY

### 2.1 Reykjanes conceptual resistivity model

In order to study the sensitivity of the CSEM and MT methods for the characterization and monitoring of high-enthalpy geothermal reservoir, we have first designed a simplified 1D resistivity model of the Reykjanes reservoir based on the existing conceptual geological models (Flovenz et al. (1985), Kristinsdottir et al. (2010), Khodayar et al. (2016)), resistivity logs and MT soundings (Karlsdottir and Vilhjalmsson (2016)). It consists in a relatively unaltered and hence resistive (100 Ohm.m) layer overlying a more conductive (1 Ohm.m) smectite-zeolite rich zone (Figure 1); then, follows a more resistive (30 Ohm.m) chlorite-epidote rich zone until supercritical conditions are met (at 4km depth in RN-15/IDDP-2 well). At this point, only a handful of studies have measured in laboratory conditions the behavior of the rock electrical resistivity but it is likely that it increases due to the drop of the brine electrical conductivity (Reinsch (2016), Nono et al. (2018)). A factor three increase of the resistivity on different Icelandic rock samples has been observed (Reinsch (2016)), most likely caused by the combination of lower viscosity reduction, thermal expansion and decrease of the dielectric constant (Kummerow and Raab (2015),Nono et al. (2018)). We therefore assumed that the chlorite-epidote rich zone in supercritical conditions is three times more resistive than the chlorite-epidote rich zone (i.e. 100 Ohm.m). Depths of the different interfaces have been defined based on the existing conceptual geological models of the Reykjanes geothermal field and well data. To study the sensitivity of the CSEM and MT methods to resistivity changes at the reservoir depth, we assumed that its resistivity drops by a factor three over a 1km thick section at 4km depth, simulating a change of geothermal fluid from supercritical to liquid due thermal cooling, as expected during the thermal stimulation of the RN-15/IDDP-2 well.

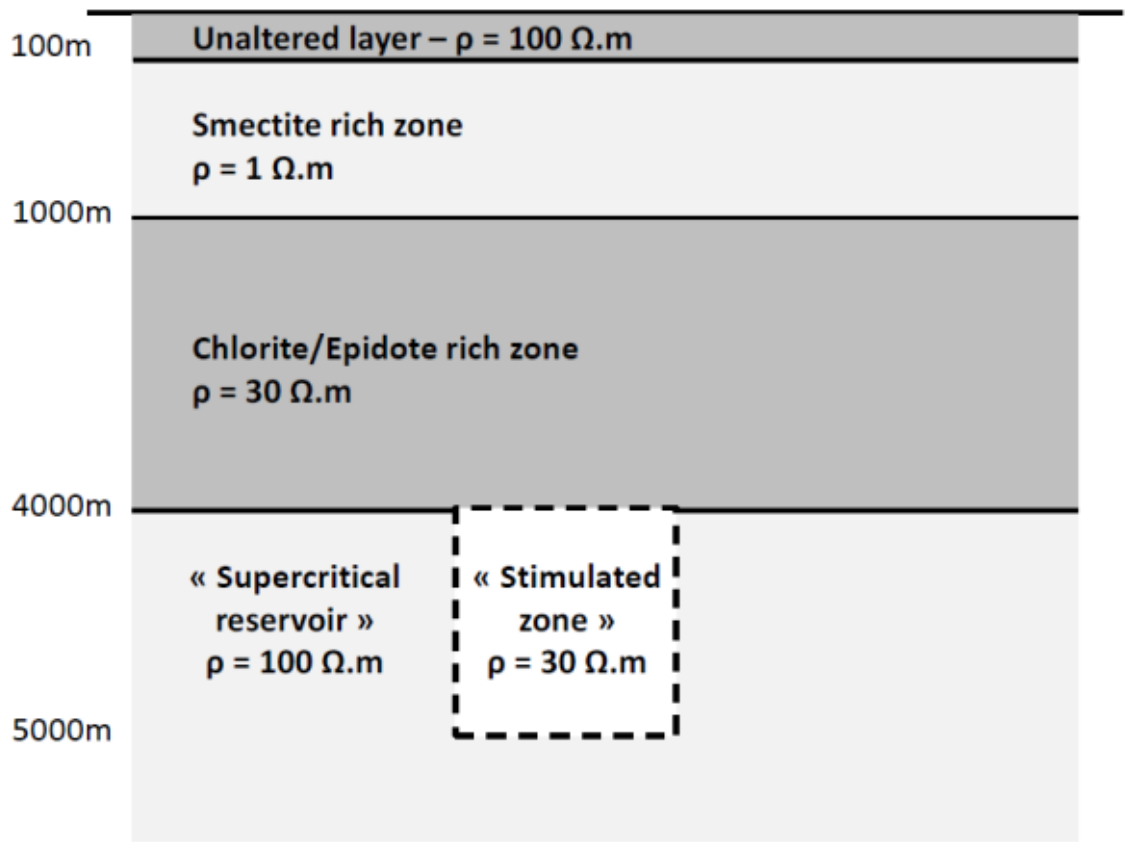

**Figure 1: Simplified resistivity model of the Reykjanes geothermal field. Depth scale is arbitrary.**

## 2.2 MT sensitivity study

We have first computed the MT impedance tensor on the aforementioned 1D resistivity models and subsequently calculated the detectability D of a time-lapse MT signal between two surveys A and B as (Ogaya et al. (2016), Thiel (2017)):

$$D^{\rho}_{MT} = |\rho_B - \rho_A| / \sqrt{\epsilon_A^2 + \epsilon_B^2}$$

where rho is the MT apparent resistivity (Ohm.m) and epsilon is the measurement error (Ohm.m). Frequencies are logarithmically distributed from 0.001Hz until 100Hz. Figure 2 displays the apparent resistivity curves for the simplified 1D resistivity model of the Reykjanes geothermal field as well as the detectability D of the signal caused by the 100 Ohm.m to 30 Ohm.m resistivity drop at 4km depth. Here, we assumed a 1% measurement error on the apparent resistivities on the base and monitor surveys (Ogaya et al. (2016)). Detectability is maximum at low frequencies (< 0.1Hz) and tops around 5.

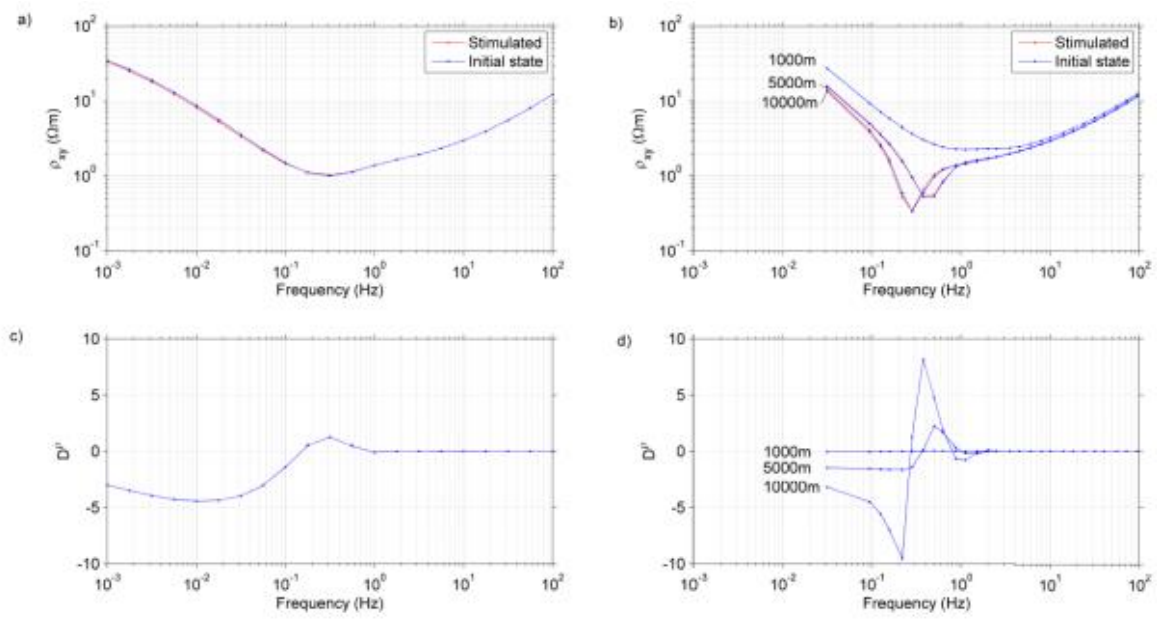

**Figure 2: a) MT and b) CSEM apparent resistivity curves for the simplified 1D resistivity model of the Reykjanes geothermal field with a 100 Ohm.m (initial state) and 30Wm (stimulated) 1km thick layer at 4km depth. c) MT and d) CSEM detectability of the 100 Ohm.m to 30 Ohm.m resistivity change over 1km at 4km depth. Offsets between the CSEM transmitter and receivers are displayed on the figure.**

## 2.3 CSEM sensitivity study

Similarly to the MT case, we have first computed the CSEM impedance tensor on the aforementioned 1D resistivity models based on the CSAMT formulation (Zonge et al. (1991)) and subsequently calculated the detectability D of a time-lapse CSEM signal between two surveys A and B as:

$$D^{\rho}_{CSEM} = |\rho_B - \rho_A| / \sqrt{\epsilon_A^2 + \epsilon_B^2}$$

where rho is the CSEM apparent resistivity (Ohm.m) and epsilon is the measurement error (Ohm.m).

CSEM fundamental frequencies range from 1/32s until 32Hz and increase by a factor 4, as typically used during CSEM field surveys (Coppo et al. (2016)). We also calculated the CSEM response for the first fourth odd harmonics of the aforementioned fundamental frequencies to obtain a well sampled spectrum from 1/32s until 100Hz. Figure 2 displays the CSEM apparent resistivity curves for the simplified 1D resistivity model of the Reykjanes geothermal _field (figure 1) as well as the CSEM detectability D of the signal caused by the 100 Ohm.m to 30 Ohm.m resistivity drop at 4km depth. Here also, we assumed a 1% error on apparent resistivities as observed on our actual measurements. Detectability is high at low frequencies (< 1Hz) and long transmitter-receiver offset (10km). Detectability tops around 10 at intermediate frequencies (0.1 - 1Hz) i.e. in the transition zone between the far and near-field CSEM response (Zonge et al. (1991)).

## 2.4 MT vs CSEM sensitivity

The CSEM and MT detectability computed on the simplified 1D resistivity model of the Rekjanes geothermal field (figure 2) shows that for a similar noise level over the whole frequency band, the sensitivity to a resistivity change within the resistive reservoir is likely to be higher on CSEM data than on MT data, most likely due to the superior sensitivity of the CSEM technique to resistors compared to MT (Constable and Weiss (2006); Weidelt (2007); Constable et al. (2009); Commer and Newman (2009)).

In addition, the use of a CSEM transmitter allows to control and hence potentially decrease the measurement error on apparent resistivities. This provides an unique opportunity to increase the detectability and hence sensitivity of the EM monitoring method to resistivity changes at the reservoir level (Siripunvaraporn et al. (2018)).

## 3. REYKJANES TIME-LAPSE EM SURVEYS

### 3.1 Time-lapse EM data acquisition

Time-lapse CSEM surveys have been acquired in September 2016, while drilling of RN-15/IDDP-2 well and in August 2017, after the thermal stimulation of the well. It used a double orthogonal horizontal electric dipole for the transmitter (figure 3), 3km north of the geothermal field providing two polarizations called POL1 (900m-long dipole between E1 and E2) and POL2 (900m-long dipole between E2 and E3). Its position is such that the mid-point of the longest transmitter-receiver offsets (7km) is located in the vicinity of the target of interest and such that injection electrodes can be installed in conductive superficial material (here, a swamp) to ensure a good electrical coupling of the transmitter with the ground. In the end, we managed to inject repeatably a current of about 30A at 560V with a Metronix TXM22 during both baseline and monitor CSEM surveys. This signal was successfully picked up by all our CSEM stations deployed over the Reykjanes peninsula (figure 3). To adequately characterize the subsurface, a broad band set of CSEM frequencies (from 1/32s up to 1024Hz) was acquired with a minimum set of 50 cycles at low-frequencies to ensure proper stacking of any random noise. The waveforms were seven square waves of fundamental frequencies ranging from 1/32s up to 128Hz increasing with a factor 4. A total number of 22 CSEM recording stations were deployed during the baseline and monitor surveys. They were Metronix ADU07 acquisition systems, MFS07 or MFS06 magnetic coils and two orthogonal 100m long electric dipoles oriented North-South and East-West. MT data have been acquired with the same equipment during the night shifts of the baseline survey i.e. when the CSEM transmitter was off. Given the results of the baseline MT survey (see section MT analysis), MT stations were only deployed a couple of hours during the monitor survey, not long enough on the ground to provide reliable low-frequency MT data. All recording equipment (electrodes, magnetometers, recording units) have been positioned with a differential GPS with a centimeter accuracy and replaced at the same position during the monitor survey to minimize positioning errors. When possible, electrodes and magnetometers have been put back in the same holes into the ground. Similarly, the transmitter electrodes and cables have been dGPS positioned and re-installed at the same position during the monitor survey.

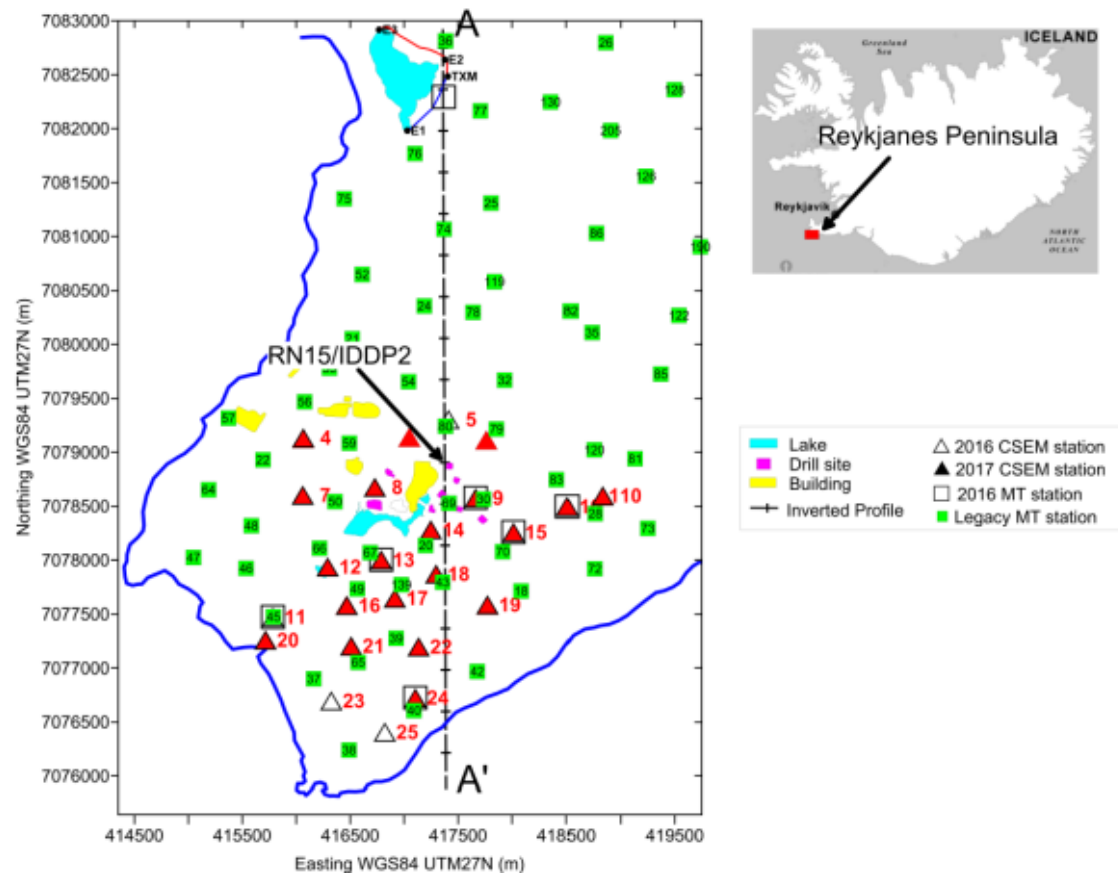

**Figure 3: Map of the Reykjanes geothermal field and time-lapse CSEM and MT survey lay-out. CSEM transmitter is labelled TXM and recording stations with numbers. Section AA' represents the axis of 2.5D inverted model.**

### 3.2 MT data analysis

Seven stations were used for MT acquisition during the baseline survey (Figure 3). Each MT station dataset consists in one hour of recordings at 4096Hz sampling frequency and at least 12 hours at 512 Hz. A distant synchronous MT station, located 80 km away, was used as a remote reference (hereafter site 100). MT sounding consistency quality assessment was performed using apparent resistivity and phase curves inter-comparison between single site and combinations of remote reference results. Phase tensor consistency analysis was performed, as advocated by Booker (2014): "Smooth variation of the phase tensor with period and position is a strong indicator of data consistency.".

In order to assess the quality of the MT data in the [1 mHz-128 Hz] band, we show the normalized phase tensor (hereafter PT), i.e the phase tensor with longer axis Phimax normalized to 1, is displayed for all frequency and RR combination. Ellipses are filled with a color bar indexed either on their their ellipticity value (left panel on figure 4) or their beta angle (right panel, same figures). Low values of ellipticity diagnose a 1D medium (Bibby et al. (2005) while beta angles absolute values below 3° diagnoses a 2D medium (Booker (2014)).

In any remote reference combination (indexed by vertical scale ticks on figure 4), discontinuous PT behavior are observed for 4 soundings (sites 9 10 11 and 24), leading to rejection of those data for interpretation. Site 13 display a smoother and coherent behavior in the high frequency (above 1 Hz) when combined with sites 10 and 100. Site 15 display smooth PT behavior at frequencies below 0.1 Hz and above 5 Hz.

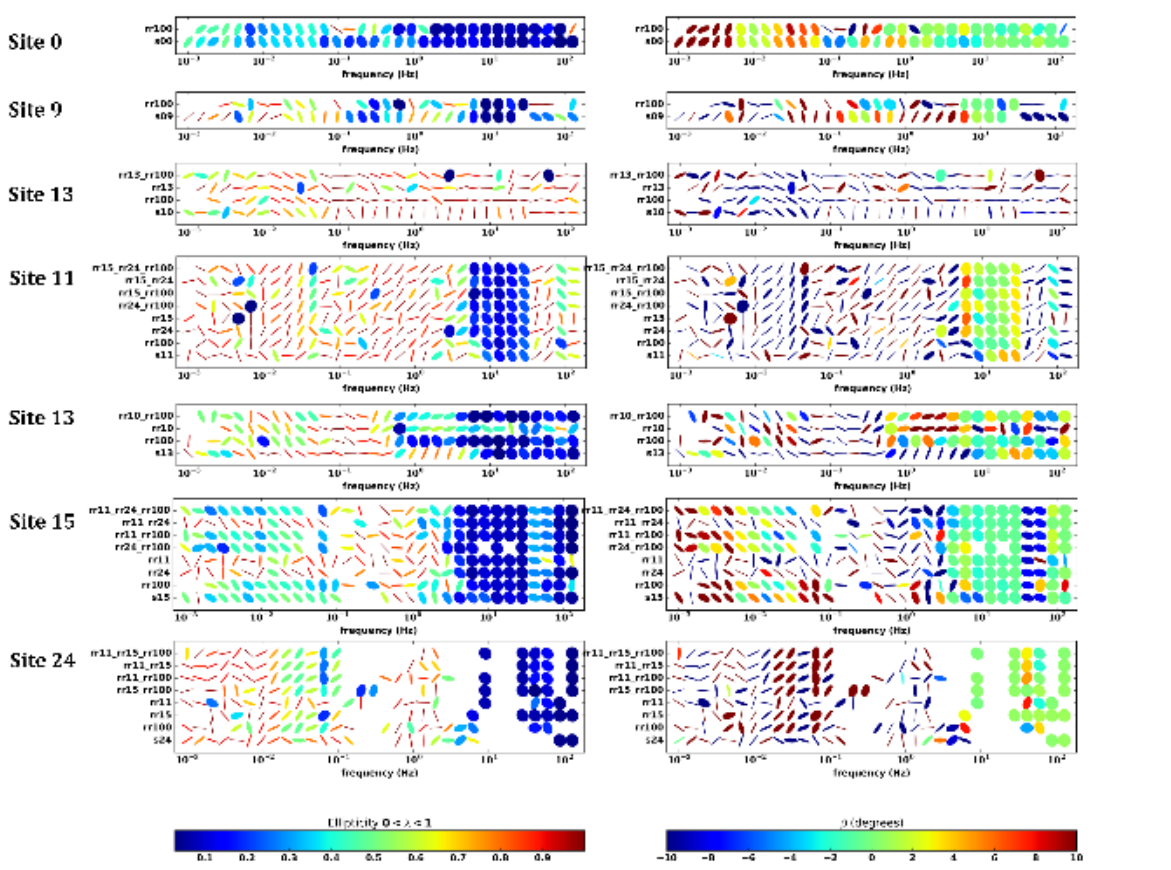

**Figure 4: Multiple remote reference two-stage bounded inuence processing results. Comparison of normalized phase tensor (PT) for each possible combination of remote reference. PT are filled with color bar indexed on their ellipticity value (left panel) and their beta angle value (right panel).**

"Best" soundings 00, 13 and 15 are displayed on figure 5 for single site (SS) processing and maximum number of RR two-stage processing. Error bars on both phase and apparent resistivity are significantly larger on multiple RR curves. Consistently with figure 4, MT soundings are inconsistent in the [0.1-5] Hz frequency band for site 15, and below 1 Hz for site 13. SS curves shows non physical apparent resistivity decreases (up to 3 order of magnitude decrease on rhoyx for site 13) in the [0.05-5] Hz band, which tends to disappear on the RR curves. Still RR curves are scattered. On site 00, MT curves are smoother in SS mode.

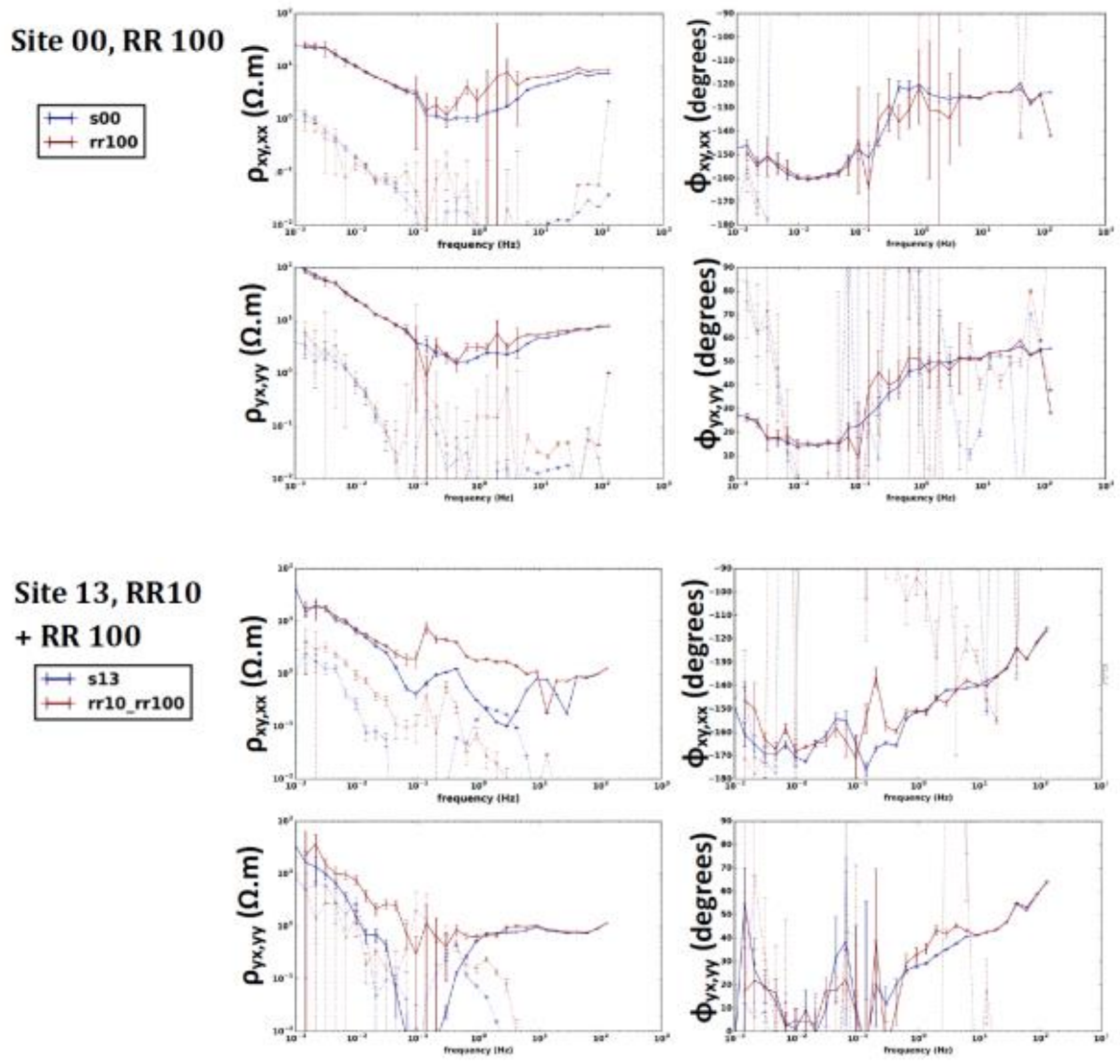

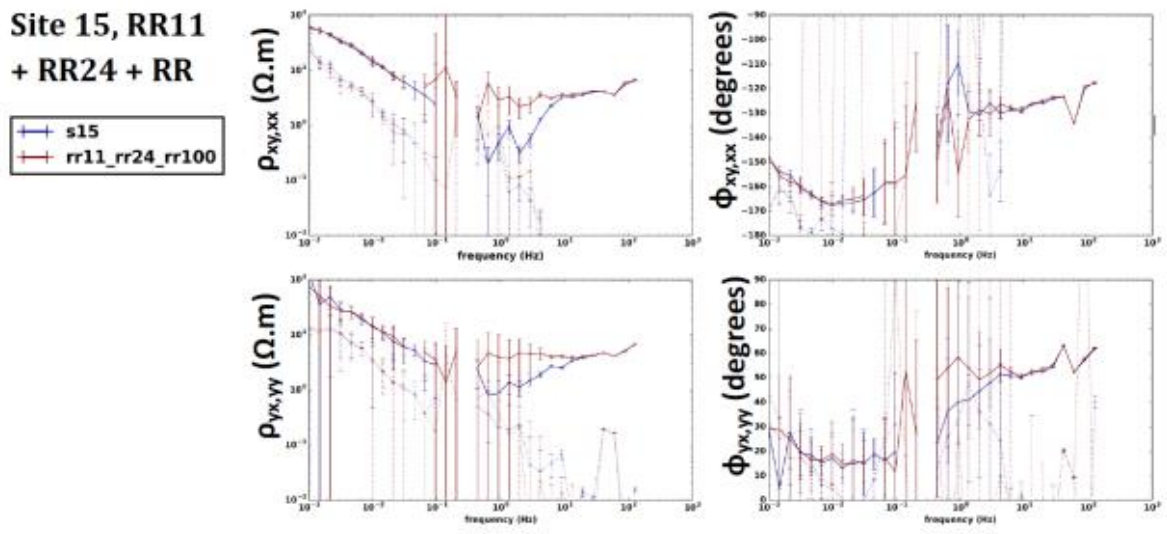

**Figure 5: MT soundings for Single Site (blue curves) and maximum number of remote reference processing for sites 00, 13 and 15 (red curves). Apparent resistivity curves rhoxy and rhoyx and phases phixy and phiyx are shown in continuous lines, components xx and yy in dashed lines.**

Due to intense anthropogenic activity in the area during RN-15/IDDP-2 drilling phase, MT soundings are of bad quality and cannot be used for interpretation. Despite the use of combinations between local and distant remote reference and bounded influence processing, a signal incoming from a near-field source persists in the data and creates either fake resistivity variations or large amplitude scatter in the frequency curves. MT imaging and subsequently monitoring with such EM noise conditions will not lead to reliable enough results. Since the baseline MT were of poor quality, we did not deem necessary to acquire MT data during the monitor time-lapse survey.

### 3.3 CSEM data analysis

To assess the CSEM survey repeatability of the time-lapse surveys, we have compared the amplitude and phase variations of the PE major axis of the horizontal electric field at station 18 between the baseline and monitor surveys (figure 6). We define the repeatability R of electric field measurements as:

$$R_{AB}{}^{E} = |E_B - E_A|/((E_A + E_B)/2)$$

where E is the amplitude of the electric field normalized by the transmitter dipolar moment (V/Am2), A and B refers to the baseline and monitor surveys, respectively. Over the whole frequency band, repeatability is within 2-3% and 2-3° for the amplitudes and phases, respectively but the presence of strong external noise on the baseline or monitor surveys on some specific frequency bands (e.g. 1/32s at low frequencies, 50Hz and harmonics at high frequencies) degrades again significantly the repeatability up to 10% and 10° on the amplitudes and phases, respectively. Although weather was humid during the baseline and dry during the monitor survey, the change of the top soil water saturation and hence resistivity seems to have a limited influence on survey repeatability.

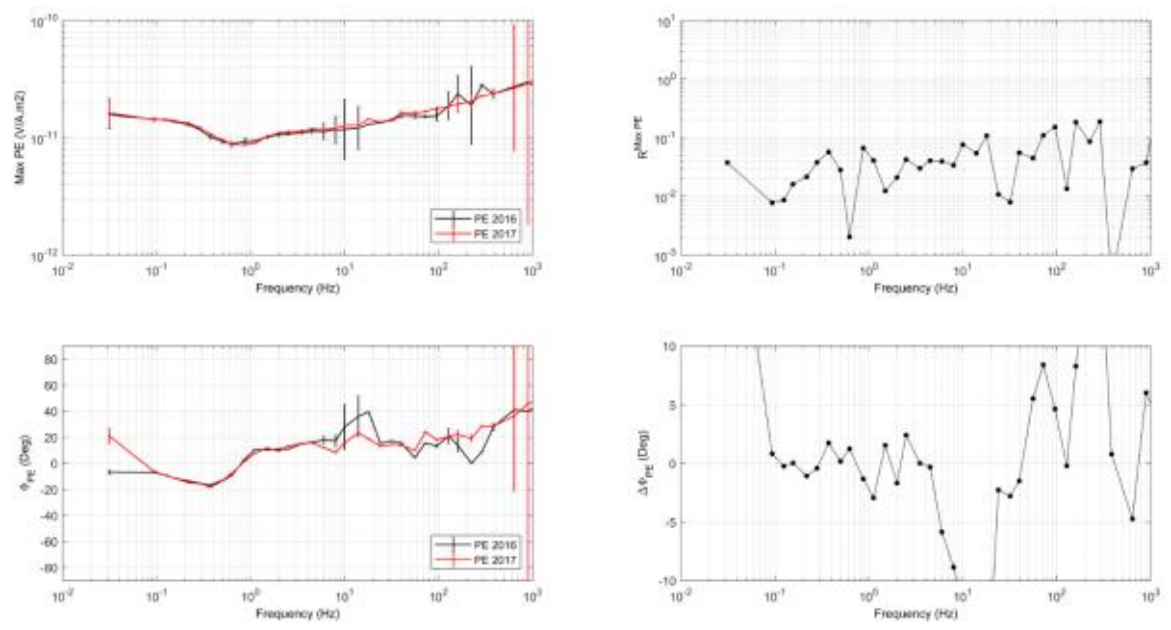

**Figure 6: Left: amplitudes (top) and phases (bottom) of the PE major axis of the horizontal electric field measured at station 18 during the baseline (black) and monitor (red) surveys as a function of the CSEM fundamental frequencies and associated first fifth odd harmonics. Noise estimates are displayed as errorbars. Right: Repeatability R of the amplitudes (top) and phase difference (bottom) of the PE major axis of the horizontal electric field between station 18 measured during baseline and monitor surveys as a function of the CSEM fundamental frequencies and associated first fifth odd harmonics.**

Interestingly, similar conclusions hold for the entire time-lapse dataset. Indeed, when comparing the repeatability of the amplitudes of the PE major axis of the horizontal electric field with the baseline and monitor signal to noise ratio (figure 7), the trend is a clear decrease of the repeatability R with increasing signal to noise ratio i.e. with decreasing level of external noise. Since the frequencies of interest for deep reservoir monitoring are low, we have limited our analysis to frequencies less than 10Hz. This observation demonstrates that for our time-lapse CSEM procedure, the signal to external noise ratio of the repeated EM measurements is the most important parameter to control in order to achieve a good survey repeatability. Contrary to MT monitoring experiments where the practitioner has limited control on the source strength and hence on the achievable survey repeatability, the CSEM signal to noise ratio can be controlled and increased at will by simply increasing the transmitter dipolar moment (e.g. longer electric dipole transmitter and/or stronger power generator) and/or recording signals for longer periods of time to increase the chance of stacking-out random external noise.

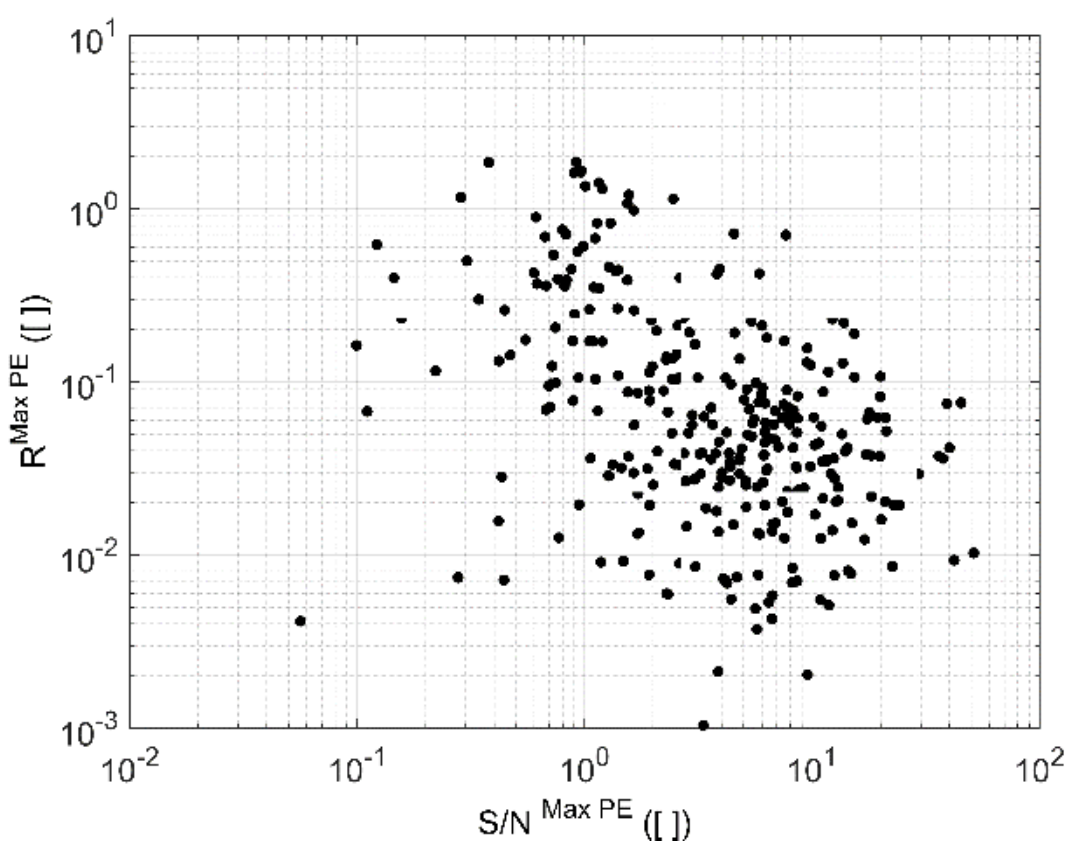

**Figure 7: Repeatability R of the amplitudes of the major PE axis of the horizontal electric fields between the baseline and monitor surveys as a function of the combined baseline and monitor signal to noise (S/N) ratios on their amplitudes. Only CSEM fundamental frequencies and associated first fifth odd harmonics less than 10Hz are displayed.**

In order to identify time-lapse signals in our dataset related to the thermal stimulation of the RN-15/IDDP2 well, we have calculated the amplitude and phase change of the polarization ellipse of the horizontal electric field between the monitor and baseline surveys (figure 8, phase not shown). We focused on the stations located along the aforementioned profile as it crosses the producing reservoir and the RN-15/IDDP-2 well. For the stations with a high signal to noise ratio and therefore high repeatability (stations 09, 14, 18, 19), no clear and consistent time-lapse anomaly related to the RN-15/IDDP-2 thermal stimulation can be identified. Indeed, observed time-lapse anomalies are random and stay within the measurement error. The only significant anomalies occur at frequencies where the signal to noise and hence repeatability is poor (e.g. 0.03125Hz, 50Hz) and are likely to be related to external sources of noise.

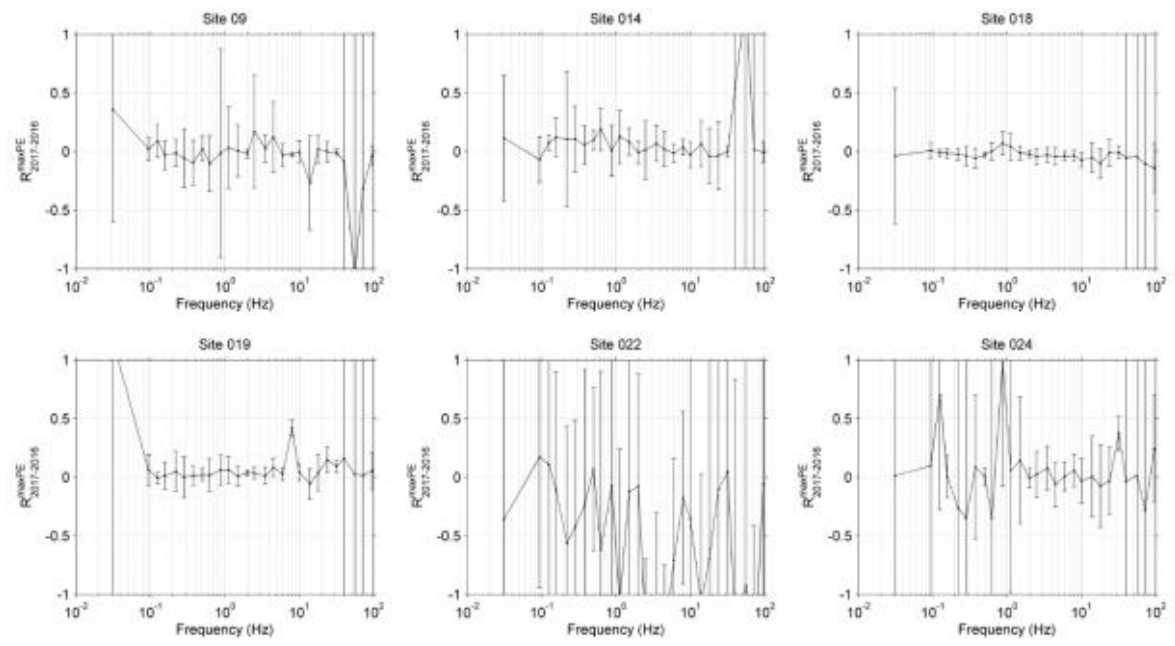

**Figure 8: Relative amplitude change of the major axis of the polarization ellipse of the horizontal electric field for the stations 09, 14, 18, 19, 22 and 24 between the monitor and baseline CSEM surveys as a function of frequency. Vertical bars indicate the estimated time-lapse amplitude measurement error.**

## 4 CSEM AND MT INVERSIONS

In this section, we have performed an inversion CSEM and MT data to confront and validate the CSEM and MT results with the logged resistivities in the RN-15/IDDP-2 well. For this calibration, we only inverted stations along a profile running from the transmitter and crossing the producing geothermal reservoir (figure 3). For the inversion, we used the 2.5D MARE2DEM inversion code (Key (2016)).

### 4.1 CSEM inversion

We inverted the amplitudes of the PE major axis of the horizontal electric field from seven CSEM stations located in the vicinity of the selected profile (stations 05, 09, 14, 18, 22 and 24). Inverted frequencies were 1/32s, 1/8s, 1/2s, 2Hz, 8Hz, 32Hz and associated first fifth odd harmonics up to 50Hz. Both POL1 and POL2 transmitter polarizations were inverted. Data from either the baseline or monitor survey were used depending on their signal to noise ratio. We limited the frequency band on the high side to 50 Hz due to the presence of strong external noise (e.g. 50Hz and harmonics, industrial noise). The starting model of the CSEM inversion was a homogeneous 2 Ohm.m half-space. Numerical simulations of the impact of the land/ocean interface showed that stations nearby the coast may be affected by the presence of the conductive sea over a large frequency band but since the area of interest (RN-15/IDDP-2 well) is located far away from the coast (at least 2km), we did not include it. Future 3D inversions will however require to include such an interface.

To assess the convergence of the inversion and quality of the data fit, we calculated RMS misfits based on measurement errors (Key (2016)). Measurement errors have been estimated from the external noise levels calculated at the processing stage. The target misfit is set to 1 and we consider the data fit to be satisfactory when misfits are small (as close as possible to unity) and have been significantly reduced during the inversion process (typically several units). Here, initial misfits were in the 10-20 range and dropped into the 2-5 range after 15 iterations, leading a satisfactory data fit over the whole frequency band (figure 9). Only station 14 has a RMS misfit great than 10, most likely due to a remaining static effect as evidenced by the similar shapes of the modelled and observed amplitude curves.

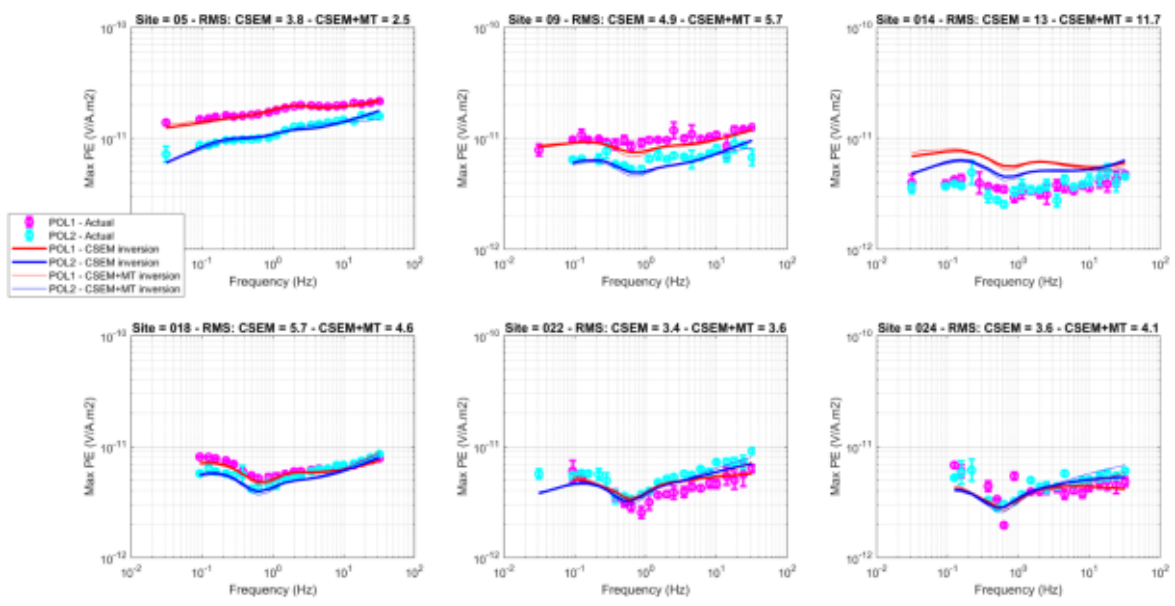

**Figure 9: Observed (dots) and modelled (solid lines) after 2.5D inversion of the amplitudes of the major axis of the polarization ellipses of the horizontal electric field as a function of the CSEM frequencies for POL1 (red/magenta) and POL2 (blue/cyan) transmitter polarizations for stations 05, 09, 14, 18, 22 and 24. Each panel corresponds to a different CSEM receiver along the inversion line. Measurement errors are displayed as vertical bars. Thin and thick solid modelled curves corresponds to the CSEM only and joint CSEM and MT inversions, respectively.**

The resulting resistivity model as well as the average resistivities logged in the RN-15/IDDP-2 well are displayed on figure 10. The shallow conductive smectite-rich caprock is well imaged in the vicinity of the RN-15/IDDP-2 well, with a resistivity (<5 ohm.m) and thickness (approximately 1200m) in good agreement with the logged values. The underlying more resistive chlorite/epidote rich layer is also imaged but deeper than 2km, the recovered resistivities are too low (20 Ohm.m vs 50/100 Ohm.m in the well). To explain this discrepancy, we have computed the Jacobian or sensitivity matrix at the last iteration of the inversion (figure 11). Higher values indicate areas where the dataset is highly sensitive to a change in resistivity. The sensitivity of the CSEM setup is clearly non-uniform with the highest sensitivity towards the mid-point between the CSEM transmitter and receiver grid (around 3km from the transmitter) i.e. in the vicinity of the RN15/IDDP-2 well (located at 3.7km distance from the transmitter), confirming that the transmitter and receiver layout is adequate for imaging resistivity variations in this area. It however also shows that the sensitivity at 4/5 km depth is low (at least two orders of magnitudes less than in the first 1.5km), possibly explaining why the resistivity values recovered from the CSEM inversion are too low compared to the logged ones. Finally, figure 11 also shows that the CSEM sensitivity is poor underneath the transmitter and the most distant receivers (distances greater than 5km from the transmitter). These low sensitivity areas explain most likely the unexpected absence of the conductive layer at large distances from the transmitter (greater than RX18) and its unexpected thickening at negative distances from the transmitter. Similarly, at shallow depth (< 1.5km) between the transmitter and first receiver (RX05), artefacts may be present due to the low sensitivity of the CSEM setup. This illustrates the difficulty of imaging complex resistivity variations with only CSEM transmitter and the need for multiple transmitter positions to obtain a more homogeneous sensitivity matrix.

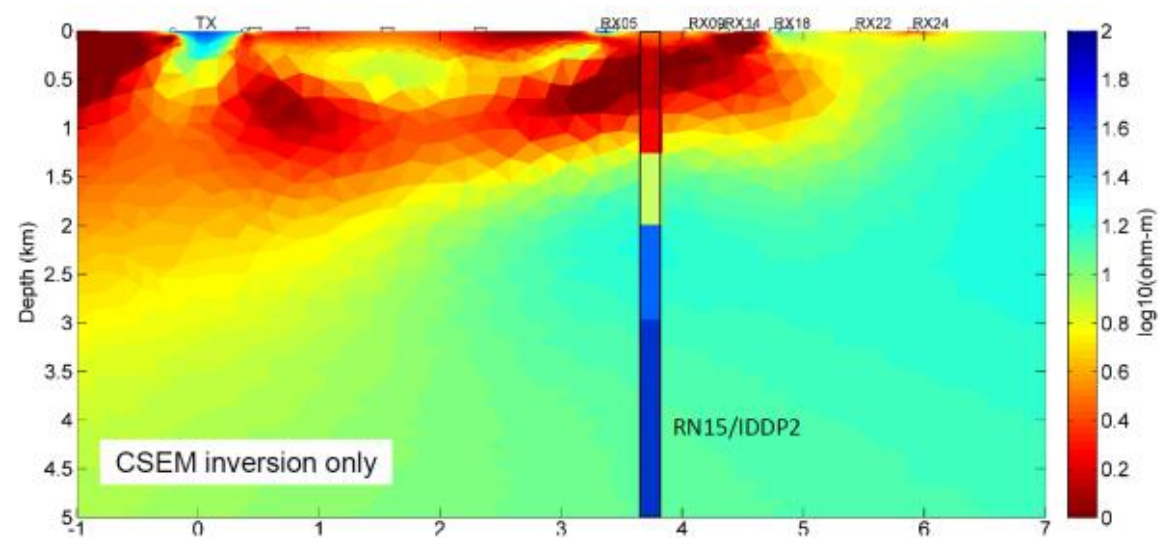

**Figure 10: Resistivity model (log scale) obtained after the 2.5D inversion of the CSEM data only from CSEM stations 05, 09, 14, 18, 22 and 24.**

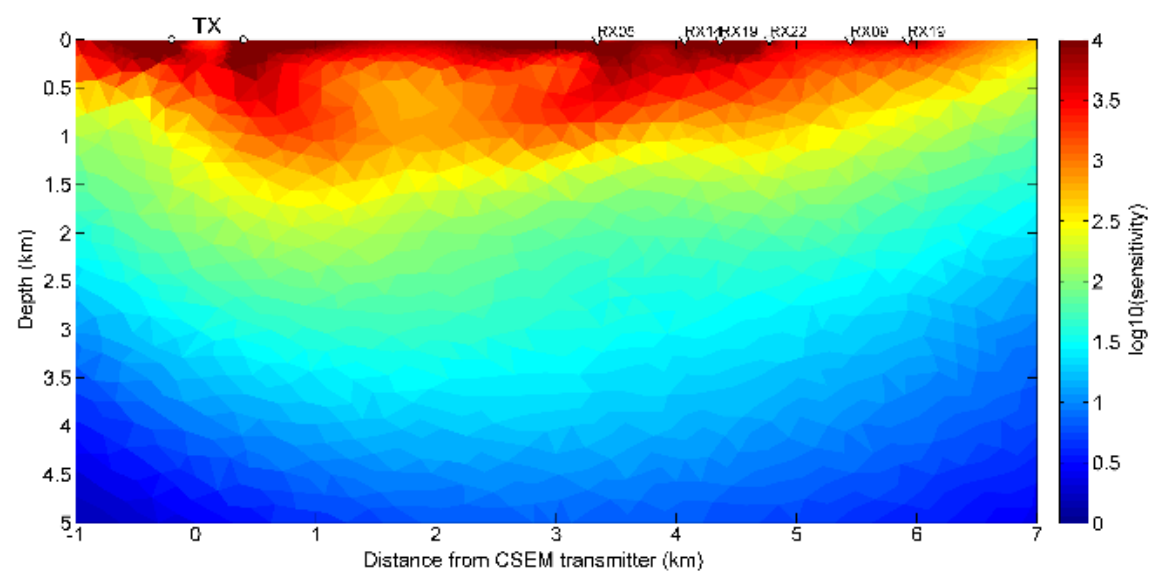

**Figure 11: Sensitivity model (Jacobian matrix in log scale) obtained after the 2.5D CSEM inversion of the stations 05, 09, 14, 18, 22 and 24.**

### 4.2 Joint MT and CSEM inversion

To compensate for the low sensitivity at depth of our CSEM setup, additional constraints (e.g. structural, petrophysical) and/or datasets (e.g. MT, resistivity logs) may be necessary (Scholl et al. (2010)). In an attempt to increase the resolution of the resistivity image at depths greater than 2/3km, we have looked into the possibility of jointly inverting CSEM and MT over the area of interest (Abubakar et al. (2011)). Since our MT dataset is of poor quality, we used the legacy MT dataset collected over the Reykjanes geothermal field instead (Karlsdottir and Vilhjalmsson (2016)).

We first inverted the apparent resistivities and phases of the non-diagonal components of the MT impedance tensor for seven MT stations nearby our CSEM stations along the profile of interest (figure 3). Frequencies range from 0.001Hz until 100Hz. Final RMS misfits are close to unity, providing a satisfactory data match (figure 12). The resulting resistivity model as well as the average resistivities logged in the RN-15/IDDP-2 well are displayed on figure 13. Here also, the shallow conductive smectite-rich caprock is well imaged with inverted resistivities (<5 Ohm.m) in good agreement with the logged values. Nevertheless, the depth of the base of this conductive layer does not match well with the well observations (a few hundreds of meters difference). Contrary to the CSEM inversion, the underlying more resistive chlorite/epidote rich layer is well imaged with highly resistive layers (up to 100 Ohm.m at 5km depth).

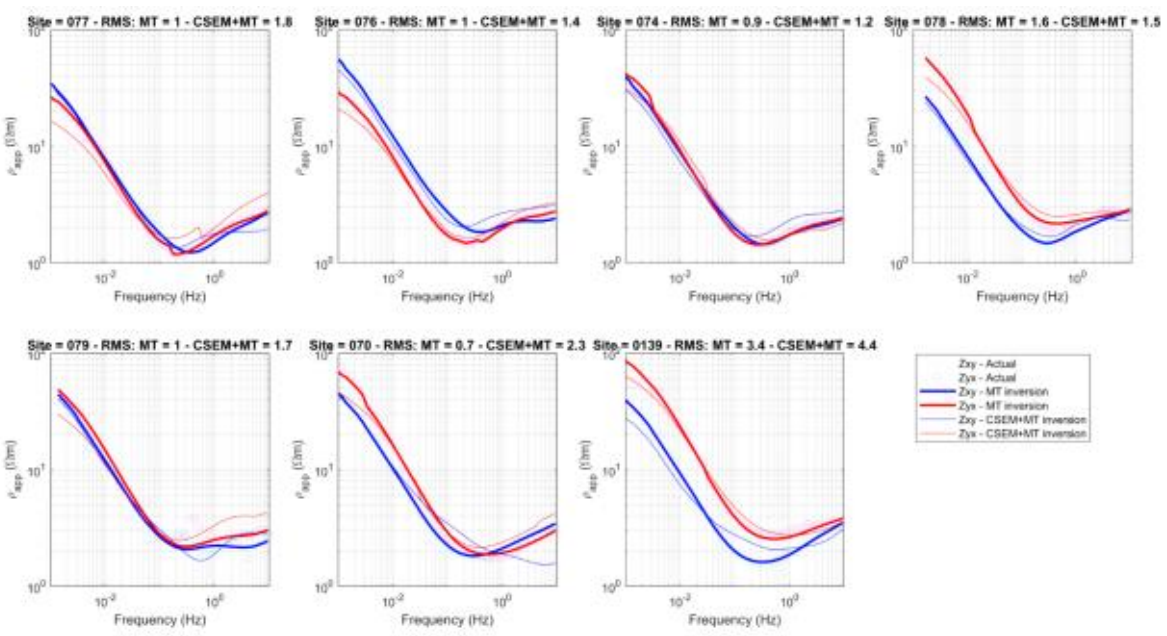

**Figure 12: Observed (dots) and modelled (solid lines) after 2.5D inversion of the amplitudes and phases of the non-diagonal components of the MT impedance tensor as a function of frequency for MT stations 77, 76, 74, 78, 79, 70 and 139. Each panel corresponds to a different MT station along the inversion line. Thin and thick solid modelled curves corresponds to the MT only and joint CSEM and MT inversions, respectively.**

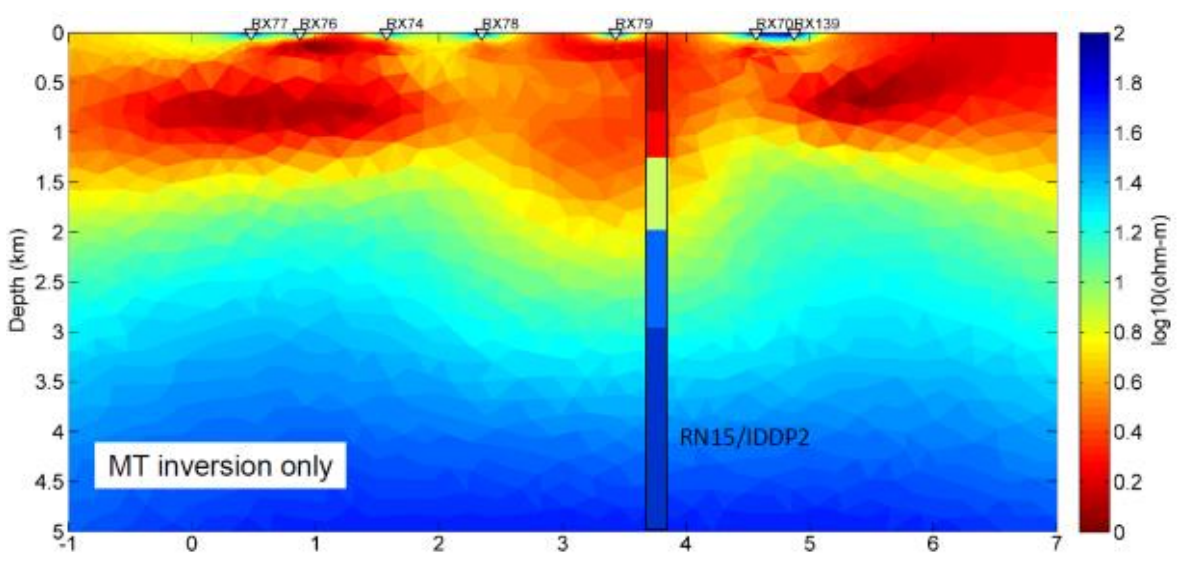

**Figure 13: Resistivity model (log scale) obtained after the 2.5D inversion of the MT data only from MT stations 77, 76, 74, 78, 79, 70 and 139.**

To take advantage of both CSEM and MT datasets, we have jointly inverted the amplitudes of the electric field for the CSEM stations 05, 09, 14, 18, 22 and 24 with the apparent resistivities and phases of the non-diagonal components of the MT impedance tensor for MT stations 77, 76, 74, 78, 79, 70 and 139. CSEM and MT data fit are displayed on figure 9 and figure 12. Overall, misfits are small and similar to the CSEM only and MT only cases, providing a satisfactory data fit. However, RMS misfits are slightly larger than the standalone cases, simply due to the fact that additional constraints have been introduced in the inversion process by the addition of new data. The resulting resistivity model as well as the average resistivities logged in the RN15/IDDP2 well are displayed on figure 14. Interestingly, both the shallow conductive smectite-rich caprock and the underlying resistive chlorite/epidote rich layer layer are now well imaged and in good agreement with the logged values. Furthermore, the depth of transition zone between the caprock and the deeper and more resistive material fits now very well with the well observations. This good match demonstrates the validity of CSEM and MT measurements for estimating and hence monitoring resistivity variations within the Reykjanes geothermal reservoir.

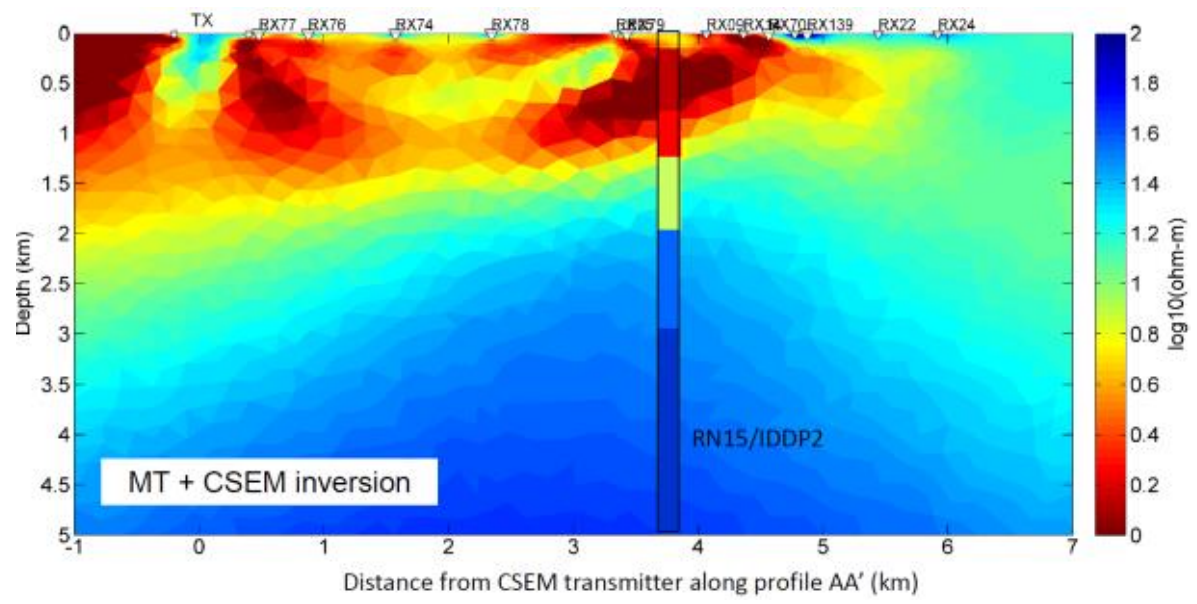

**Figure 14: Resistivity model (log scale) obtained after the joint 2.5D inversion of CSEM data from CSEM stations 05, 09, 14, 18, 22 and 24 and MT data from MT stations 77, 76, 74, 78, 79, 70 and 139.Resistivity model (log scale) obtained after the 2.5D inversion of the MT data only from MT stations 77, 76, 74, 78, 79, 70 and 139.**

## 5 DISCUSSION

Despite the high degree repeatability of the CSEM measurements between the Reykjanes baseline and monitor (a few percent on the amplitude of the electric field), no clear and consistent time-lapse anomaly related to the RN-15/IDDP-2 thermal stimulation has been identified. A most likely explanation is related to the weakness of the time-lapse CSEM signals in comparison to the achieved repeatability. To demonstrate this, we have calculated the detectability of time-lapse CSEM signals based on electric field amplitudes as a function of the size of the stimulated zone (here, width) and measurements errors for the 2.5D Reykjanes conceptual model (figure 15). It clearly shows that the amplitude of the time-lapse signal is strongly related to the volume of the stimulated area (here, its width as its height is kept fixed at 1km). For the repeatability achieved during the actual Reykjanes time-lapse survey (a few percent), it indicates that a time-lapse signal can be observed (D greater than 1) only if the stimulated area is larger than 500m in width. During the drilling of RN-15/IDDP-2 well, high-permeability circulation-fluid loss zones were detected below 3 km depth to bottom. The largest one occurred at around 3.4 km depth with permeable zones encountered below 3.4 km accepting less than 5%percent of the injected water. It is therefore likely that most of the fluid injected during the thermal stimulation leaked into this zone between 3 and 3.4km depth. Since the total volume of injected cold water was roughly 100000m3 in one month and the porosity of the in-situ rock is low (a few percent), the lateral extent of the stimulated zone does not exceed a couple of hundreds meters and most likely well below the detectability threshold achieved for our actual survey. To pick such a small signal up, even more repeatable measurements would be required (less than a percent, figure 15) or alternatively, a more sensitive CSEM layout would need to be deployed (e.g. a borehole to surface CSEM configuration, Tietze et al. (2015)). Our CSEM time-lapse analysis is based on the amplitude and phase of the electric phase measurements but it is possible that other parameters are actually more sensitive to resistivity changes than the raw electric field measurements, like the distortion (Rees et al. (2016)) or phase tensor (Booker (2014)). The computation of such parameters have however to be adapted to the CSEM case.

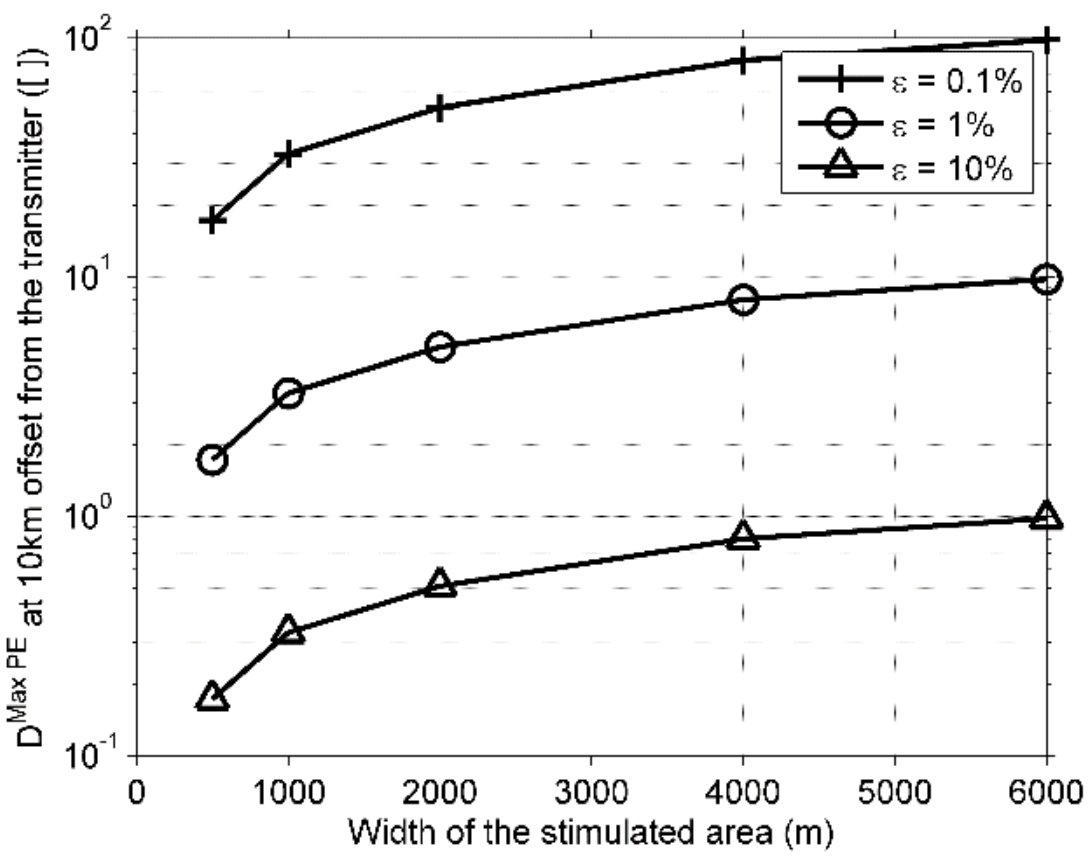

**Figure 15: Detectability of the CSEM time-lapse signal based on the amplitudes of the electric field at 10km offset from the transmitter as function of the width of the stimulated area and measurement errors for the 2.5D Reykjanes resistivity model (figure 1). Measurement errors are expressed as a percentage of the total electric field.**

## CONCLUSION

The resistivity structure of the Reykjanes geothermal field (conductive caprock overlying a more resistive high temperature reservoir) is very generic for any high-enthalpy geothermal reservoirs (Flovenz et al. (1985), Kristinsdottir et al. (2010), Khodayar et al. (2016)) and conclusions drawn on this particular example are therefore applicable to many other geothermal fields.

The CSEM calibration survey performed here shows such data provides reliable data for the imaging and monitoring of high-enthalpy geothermal reservoir. The main benefit relies on the high signal to noise ratios that can be achieved despite the presence of high levels of cultural noise. At this stage, the main drawback is caused by the limited depth of penetration (2-3km depth), most likely caused by the combination of a thick conductive and hence attenuating caprock, and the limited dipolar moment of the transmitter. Greater depths of penetration can surely be achieved using more powerful transmitters (e.g. longer dipole, higher currents) as recently developed for offshore CSEM systems (Hanssen et al. (2017)). In addition, the resistivity of the overburden has to be taken into account as more resistive overburdens can often lead to greater depths of penetration (3-4km) with similar CSEM systems (Coppo et al. (2016)).

As shown on the Reykjanes example, MT data provides a good alternative to increase the depth of investigation (> 2-3km) when CSEM data is of limited use. For monitoring purposes, the challenge is however to obtain a highly repeatable MT dataset (Abdelfettah et al. (2018)). Continuous MT and CSEM monitoring surely provides a good way to control the quality of the measurements by correlating them with subsurface phenomena but it also represents a huge logistical challenge for long term monitoring. Indeed, numerical simulations (figure 15, Orange et al. (2009), Wirianto et al. (2010), Thiel (2017)) show that only resistivity changes happening over a significant reservoir volume (e.g. after long periods of fluid injection/production) may lead to detectable EM signals. Time-lapse MT measurements alleviate this logistical constraint but as shown with our particular example, significant efforts have to be made to ensure suffcient data quality during both baseline and monitor MT surveys, especially when performed in highly industrialized areas with high levels of electromagnetic noise.

## Acknowledgements

The IDDP-2 was funded by HS Orka, Landsvirkjun, Orkuveita Reykjavikur, and the National Energy Authority in Iceland, together with Statoil, the Norwegian oil and gas company. The IDDP-2 has also received funding from the DEEPEGS project, European Union's HORIZON 2020 research and innovation program under grant agreement No 690771. We would like to thank HS-ORKA for providing access to the MT dataset over the Reykjanes geothermal field.

## REFERENCES

Abubakar, A., Li, M., Pan, G., Liu, J., Habashy, T., 2011. Joint mt and csem data inversion using a multiplicative cost function approach. Geophysics 76, F203-F214.

Bibby, H., Caldwell, T., Brown, C., 2005. Determinable and non-determinable parameters of galvanic distortion in magnetotellurics. Geophysical Journal International 163, 915-930.

Booker, J.R., 2014. The magnetotelluric phase tensor: a critical review. Surveys in Geophysics 35, 7-40.

Commer, M., Newman, G.A., 2009. Three-dimensional controlled-source electromagnetic and magnetotelluric joint inversion. Geophysical Journal International 178, 1305-1316.

Constable, S., Key, K., Lewis, L., 2009. Mapping offshore sedimentary structure using electromagnetic methods and terrain e_ects in marine magnetotelluric data. Geophysical Journal International 176, 431-442.

Constable, S., Weiss, C.J., 2006. Mapping thin resistors and hydrocarbons with marine em methods: Insights from 1d modeling. Geophysics 71, G43-G51.

Coppo, N., Darnet, M., Harcouet-Menou, V., Wawrzyniak, P., Manzella, A., Bretaudeau, F., Romano, G., Lagrou, D., Girard, J.F., 2016. Characterization of deep geothermal energy resources in low enthalpy sedimentary basins in belgium using electro-magnetic methods-csem and mt results, in: European Geothermal Congress 2016.

Flovenz, O.G., Georgsson, L.S., Arnason, K., 1985. Resistivity structure of the upper crust in iceland. Journal of Geophysical Research: Solid Earth 90, 10136-10150.

Hanssen, P., Nguyen, A.K., Fogelin, L.T., Jensen, H.R., Skar, M., Mittet, R., Rosenquist, M., Suilleabhain, L. O., van der Sman, P., 2017. The next generation offshore csem acquisition system , 1194-1198.

Karlsdottir, R., Vilhjalmsson, A.M., 2016. Sandvik, reykjanes peninsula. 3d inversion of mt data. H2020 DEEPEGS report _ISOR-2016/041.

Key, K., 2016. Mare2dem: a 2-d inversion code for controlled-source electromagnetic and magnetotelluric data. Geophysical Journal International 207, 571-588.

Khodayar, M., Nielsson, S., Hickson C., Gudnason, E., Hardarson, B., Gudmundsdottir, V., Halldorsdottir, S., Oskarsson, F., Weisenberger, T., Bjornsson, S., 2016. The 2016 conceptual model of reykjanes geothermal system, sw iceland. DEEPEGS Report ISOR-2016/072.

Kristinsdottir, L.H., Flovenz, O.G., Arnason, K., Bruhn, D., Milsch, H., Spangenberg, E., Kulenkamp, J., 2010. Electrical conductivity and p-wave velocity in rock samples from high-temperature Icelandic geothermal fields. Geothermics 39, 94-105.

Kummerow, J., Raab, S., 2015. Temperature dependence of electrical resistivity-part i: Experimental investigations of hydrothermal uids. Energy Procedia 76, 240-246.

Nono, F., Gibert, B., Parat, F., Loggia, D., Cichy, S.B., Violay, M., 2018. Electrical conductivity of Icelandic deep geothermal reservoirs up to supercritical conditions: Insight from laboratory experiments. Journal of Volcanology and Geothermal Research.

Ogaya, X., Ledo, J., Queralt, P., Jones, A.G., Marcuello, A., 2016. A layer stripping approach for monitoring resistivity variations using surface magnetotelluric responses. Journal of Applied Geophysics 132, 100-115.

Orange, A., Key, K., Constable, S., 2009. The feasibility of reservoir monitoring using time-lapse marine csem. Geophysics 74, F21-F29.

Rees, N., Heinson, G., Krieger, L., 2016. Magnetotelluric monitoring of coal seam gas depressurization. Geophysics 81, E423-E432.

Reinsch, T., 2016. Physical properties of rock at reservoir conditions. FP7 IMAGE Report .

Scholl, C., Hallinan, S., Miorelli, F., Watts, M., 2010. Geological consistency from inversions of geophysical data. 79th EAGE Conference and Exhibition 2017.

Siripunvaraporn, W., Bedrosian, P.A., Li, Y., Patro, P.K., Spitzer, K., Toh, H., 2018. Special issue studies on electromagnetic induction in the earth: recent advances". Earth, Planets and Space 70, 47.

Thiel, S., 2017. Electromagnetic monitoring of hydraulic fracturing: Relationship to permeability, seismicity, and stress. Surveys in Geophysics , 1-37.

Tietze, K., Ritter, O., Veeken, P., 2015. Controlled-source electromagnetic monitoring of reservoir oil saturation using a novel borehole-to-surface configuration. Geophysical Prospecting 63, 1468-1490.

Weidelt, P., 2007. Guided waves in marine csem. Geophysical Journal International 171, 153-176.

Wirianto, M., Mulder, W., Slob, E., 2010. A feasibility study of land csem reservoir monitoring in a complex 3-d model. Geophysical Journal International 181, 741-755.

Zonge, K.L., Hughes, L.J., Nabighian, M., 1991. Electromagnetic methods in applied geophysics. Society of Exploration Geophysicists, SEG Electromagnetic methods in applied geophysics.